\documentclass[prl,twocolumn,showpacs,showkeys,superscriptaddress,preprintnumbers,floatfix]{revtex4-1}

\usepackage{etex}
\usepackage{ifpdf}
\usepackage{hyperref}
\usepackage{dcolumn}
\usepackage{url}
\usepackage{amsmath}
\usepackage{amscd}
\usepackage{amsfonts}
\usepackage{amssymb}
\usepackage{bm}   
\usepackage{bbm}
\usepackage{verbatim}
\usepackage{stmaryrd}
\usepackage{amsthm}
\usepackage{xcolor}
\usepackage{setspace}
\usepackage{braket}

\usepackage{tikz}
\usetikzlibrary{arrows}
\usetikzlibrary{automata}
\usetikzlibrary{decorations.pathreplacing}
\usetikzlibrary{positioning}
\usetikzlibrary{plotmarks}
\usetikzlibrary{calc}
\usetikzlibrary{patterns}
\usetikzlibrary{external}
\usepackage{pgfplots}
\pgfplotsset{compat=newest}

\usepackage{graphicx}

\theoremstyle{plain}    
\theoremstyle{plain}    
\theoremstyle{plain}    
\theoremstyle{plain}    
\theoremstyle{plain}    
\theoremstyle{plain}    
\theoremstyle{plain}    
\theoremstyle{plain}    
\theoremstyle{plain}    
\theoremstyle{plain}    
\theoremstyle{plain}    
\theoremstyle{plain}

\newcommand{\eM}     {\mbox{$\epsilon$-machine}}
\newcommand{\eMs}    {\mbox{$\epsilon$-machines}}

\newcommand{\CausalState}   { \mathcal{S} }

\newcommand{\causalstate}   { \sigma }

\newcommand{\AlternateState}    { \mathcal{R} }

\newcommand{\alternatestate}    { \rho }

\newcommand{\Cmu}       {C_\mu}
\newcommand{\hmu}       {h_\mu}
\newcommand{\EE}        {{\bf E}}

\newcommand{\PC}        {\chi}

\newcommand{\forward}{+}
\newcommand{\reverse}{-}
\newcommand{\forwardreverse}{\pm} 

\newcommand{\FutureCausalState} { {\CausalState}^{\forward} }

\newcommand{\PastCausalState}   { {\CausalState}^{\reverse} }

\newcommand{\lastindex}[2]{
  \edef\tempa{0}
  \edef\tempb{#2}
  \ifx\tempa\tempb
    \edef\tempc{#1}
  \else
    \edef\tempa{0}
    \edef\tempb{#1}
    \ifx\tempa\tempb
      \edef\tempc{#2}
    \else
      \edef\tempc{#1+#2}
    \fi
  \fi
  \tempc
}

\newcommand{\I}{\mathbf{I}}

\newcommand{\CSjoint}[1][,]{
   \edef\tempa{:}
   \edef\tempb{#1}
   \ifx\tempa\tempb
      \ensuremath{\FutureCausalState\!#1\PastCausalState}
   \else
      \ensuremath{\FutureCausalState#1\PastCausalState}
   \fi
}

\newif\ifpm
\edef\tempa{\forwardreverse}
\edef\tempb{\pm}
\ifx\tempa\tempb
   \pmfalse
\else
   \pmtrue
\fi

\renewcommand{\H}{\operatorname{H}}
\renewcommand{\I}{\operatorname{I}}

\newcommand{\kB}{k_\text{B}}  

\begin{document}

\title{Transient Dissipation and Structural Costs of\\
Physical Information Transduction}

\author{Alexander B. Boyd}
\email{abboyd@ucdavis.edu}
\affiliation{Complexity Sciences Center and Physics Department,
University of California at Davis, One Shields Avenue, Davis, CA 95616}

\author{Dibyendu Mandal}
\email{dibyendu.mandal@berkeley.edu}
\affiliation{Department of Physics, University of California, Berkeley, CA
94720, U.S.A.}

\author{Paul M. Riechers}
\email{pmriechers@ucdavis.edu}
\affiliation{Complexity Sciences Center and Physics Department,
University of California at Davis, One Shields Avenue, Davis, CA 95616}

\author{James P. Crutchfield}
\email{chaos@ucdavis.edu}
\affiliation{Complexity Sciences Center and Physics Department,
University of California at Davis, One Shields Avenue, Davis, CA 95616}

\date{\today}
\bibliographystyle{unsrt}

\begin{abstract}
A central result that arose in applying information theory to the stochastic
thermodynamics of nonlinear dynamical systems is the Information-Processing
Second Law (IPSL): the physical entropy of the universe can decrease if
compensated by the Shannon-Kolmogorov-Sinai entropy change of appropriate
information-carrying degrees of freedom. In particular, the asymptotic-rate
IPSL precisely delineates the thermodynamic functioning of autonomous
Maxwellian demons and information engines. How do these systems begin to
function as engines, Landauer erasers, and error correctors? Here, we identify
a minimal, inescapable transient dissipation engendered by physical information
processing not captured by asymptotic rates, but critical to adaptive
thermodynamic processes such as found in biological systems. A component of
transient dissipation, we also identify an implementation-dependent cost that
varies from one physical substrate to another for the same information
processing task.  Applying these results to producing structured patterns from
a structureless information reservoir, we show that ``retrodictive'' generators
achieve the minimal costs. The results establish the thermodynamic toll imposed
by a physical system's structure as it comes to optimally transduce information.
\end{abstract}


\preprint{Santa Fe Institute Working Paper 16-12-XXX}
\preprint{arxiv.org:1612.XXXXX [cond-mat.stat-mech]}

\maketitle


\setstretch{1.1}

\paragraph{Introduction} Classical thermodynamics and statistical mechanics appeal to various
reservoirs---reservoirs of heat, work, particles, and chemical species---each
characterized by unique, idealized thermodynamic properties. A heat reservoir,
for example, corresponds to a physical system with a large specific heat and
short equilibration time. A work reservoir accepts or gives up energy
without a change in entropy. Arising naturally in recent analyses of Maxwellian
demons and information engines~\cite{Maru2009, Saga2010, Horo2010, Mand012a, Mand2013, Stra2013, Bara2013, Hopp2014, Lu14a, Boyd14b, Chap15a, Merh15a, Um2015, Kosk15, Parr2015, Rana2016, Boyd16c}, \emph{information reservoirs} have come to play a
central role as idealized physical systems that exchange information but not
energy \cite{Deff2013, Saga13, Bara2014a}. Their inclusion led rather directly to an extended Second Law of
Thermodynamics for complex systems: The total physical (Clausius) entropy of
the universe and the Shannon entropy of its information reservoirs cannot
decrease in time \cite{Seif2005, Mand012a, Deff2013, Benn82, Land61a}. We refer to this generalization as the Information Processing
Second Law (IPSL) \cite{Boyd16b}.
	
A specific realization of an information reservoir is a tape of symbols where
information is encoded in the symbols' values \footnote{See
Supplementary Materials: Information reservoir implementations.}. 
To understand the role that information processing plays in the efficiencies
and bounds on thermodynamic transformations the following device has been
explored in detail: a ``ratchet" slides along a tape and interacts with one
symbol at a time in presence of heat and work reservoirs \cite{Boyd15a}. By
increasing the tape's Shannon entropy, the ratchet can steadily transfer energy
from the heat to the work reservoirs \cite{Mand012a}. This violates the
conventional formulation of the Second Law of Thermodynamics but is permitted
by the IPSL.

Since the ratchet transforms information encoded in the tape, we refer to it as
an \emph{information transducer}. Recent models of autonomous Maxwellian demons
and information engines are specific examples of information transducers. From
an information-theoretic viewpoint, these transducers are memoryful
communication channels from input to output symbol sequences~\cite{Barn13a}.
Information transducers are also similar to Turing machines in design~\cite{Stra15a}, except
that a Turing machine need not move unidirectionally. More importantly, an
information transducer is a physical thermodynamic system and so is typically
stochastic \footnote{See Supplementary Materials: Turing machines versus transducers.}. Despite this
difference, like a Turing machine a transducer can perform any computation, if
allowed any number of internal states.

Previous analyses of the thermodynamic resources required for information
processing largely focused on the minimal asymptotic entropy production
\emph{rate} for a given information transduction; see
Eq.~(\ref{eq:EntropyRateBound}) below. The minimal rate is completely specified
by the information transduction; there is no mention of any cost due to the
transducer itself. In contrast, this Letter first derives an exact expression
for the minimal \emph{transient} entropy production required for information
transduction; see Eq.~(\ref{eq:Main}). This transient dissipation is the cost
incurred by a system as it adapts to its environment. It is related to the
excess heat in transitions between nonequilibrium steady states~\cite{Hatano01,
Mand16, Riech_ftdness}. Moreover, hidden in this minimal transient dissipation,
we identify the minimal cost associated with the transducer's construction;
Eq.~(\ref{eq:CostImpl}) below. Among all possible constructions that support a
given computational task, there is a minimal, finite cost due to the physical
\emph{implementation}.

The Letter goes on to consider the specific case of structured pattern
generation from a structureless information reservoir---a tape of independent
and identically distributed (IID) symbols.  While the transducer formalism for
information ratchets naturally includes inputs with temporal structure, most
theory so far has considered structureless inputs \cite{Garn15, Boyd15a,
Mand012a, Mand2013, Bara2013, Bara2014b}. This task requires designing a
transducer that reads a tape of IID symbols as its input and outputs a target
pattern. Employing the algebra of Shannon measures \cite{Yeun08a} and the
structure-analysis tools of computational mechanics \cite{Crut12a}, we show
that the minimum implementation-dependent cost is determined by the mutual
information between the transducer and the output's ``past"---that portion of
the output tape already generated. The result is that a maximally efficient
implementation is achieved with a ``retrodictive" model of the structured
pattern transducer. Since the retrodictor's states depend only on the output
future, it only contains as much information about the output's past as is
required to generate the future. As a result it has a minimal cost proportional
to the tape's \emph{excess entropy} \cite{Crut12a}. Such thermodynamic costs
affect information processing in physical and biological systems that undergo
finite-time transient processes when adapting to a complex environment.

\paragraph{Information Processing Second Law} 
Consider a discrete-time Markov process involving the transducer's current
state $X_N$ and the current state of the information reservoir $\mathbf{Y}_N$ it processes. The latter
is a semi-infinite chain of variables over the set $\mathcal{Y}$ that the
transducer processes sequentially. $Y_N$ is the $N$th tape element, if the
transducer has not yet processed that symbol; it is denoted $Y'_N$, if the
transducer has. We call $Y_N$ an input and $Y'_N$ an output. The current tape
$\mathbf{Y}_N=Y'_{0:N}Y_{N:\infty}$ concatenates the input tape $Y_{N:
\infty}=Y_N Y_{N+1} Y_{N+2} \hdots$ and output tape $Y^\prime_{0:N}=Y'_0Y'_1
\hdots Y'_{N-2} Y'_{N-1}$. The information ratchet performs a computation by
steadily transducing the input tape process $\Pr(Y_{0:\infty})$ into the output
tape process $\Pr(Y'_{0:\infty})$.

The IPSL sets a bound on the average heat dissipation $Q_{0 \rightarrow N}$
into the thermal reservoir over the time interval $t \in [0, N]$ in terms of
the change in state uncertainty of the information ratchet and information
reservoir \cite[App. A]{Boyd15a}:
\begin{align}
\frac{\langle Q_{0 \rightarrow N}\rangle }{k_{\text B} T \ln2}
  & \geq \H[X_0, \mathbf{Y}_0]- \H[X_N, \mathbf{Y}_N] ,
\label{eq:FullBound}
\end{align} 
where $\kB$ is Boltzmann's constant, $T$ the absolute temperature of the
reservoirs, $\H[Z]$ the Shannon (information) entropy of the random variable
$Z$.

To date, studies of such information engines developed the IPSL's asymptotic-rate form:
\begin{align}
\lim_{N \rightarrow \infty} \frac{1}{N}
  \frac{\langle Q_{0 \rightarrow N}\rangle}{\kB T \ln2}
  \geq - (\hmu^\prime  - \hmu)
  ~,
\label{eq:EntropyRateBound}
\end{align}
where $\hmu^\prime$ ($\hmu$) is the Shannon entropy rate of the output (input)
tape \footnote{See Supplementary Materials: Information
generation in physical systems.} and, in addition, we assume the transducer has
a finite number of states~\cite{Boyd15a}.

The asymptotic IPSL in Eq.~\eqref{eq:EntropyRateBound} says that thermal
fluctuations from the environment can be rectified to either perform work or
refrigerate (on average) at the cost of randomizing the information reservoir
($\langle Q \rangle < 0$ when $\hmu^\prime  > \hmu$). Conversely, an
information reservoir can be refueled or `charged' back to a clean slate by
erasing its Shannon-entropic information content at the cost of emitting heat.

A crucial lesson in the physics of information is that
Eq.~(\ref{eq:EntropyRateBound}) takes into account \emph{all orders} of
temporal correlations present in the input tape as well as all orders of
correlation that the transducer develops in the output tape. An approximation
of Eq.~\eqref{eq:EntropyRateBound}, based on the inclusion of only lowest-order
(individual symbol) statistics, had been used to interpret the thermodynamic functioning of the
original models of autonomous Maxwellian Demons \cite{Mand012a,Mand2013}.
Later, Eq.~(\ref{eq:EntropyRateBound}) itself was used to identify a region in
an engine's phase diagram that is wrongly characterized as functionally useless
by the approximation, but actually is a fully functional eraser. In turn, this
motivated the construction of an explicit mechanism by which temporal
correlations in the input sequence can be exploited as a thermodynamic resource
\cite{Boyd15a}. Equation~(\ref{eq:EntropyRateBound}) also led to (i) a general
thermodynamic framework for memory in sequences and in transducers and (ii) a
thermodynamic instantiation of Asbhy's law of requisite variety---a cybernetic
principle of adaptation \cite{Boyd16b}.

Equation~(\ref{eq:EntropyRateBound}), however, does not account for
correlations between input and output tapes nor those that arise between the
transducer and the input and output. As we now show, doing so leads directly to
predictions about the relative effectiveness of transducers that perform the
same information processing on a given input, but employ different physical
implementations; cf. \cite{Garn15}. Subtracting the IPSL's asymptotic-rate version
(Eq.~(\ref{eq:EntropyRateBound})) from the IPSL's original
(Eq.~(\ref{eq:FullBound})) leads to a lower bound on the \textit{transient}
thermodynamic cost $\langle Q^\text{tran} \rangle$ of information transduction,
the Letter's central result:
\begin{align}
\label{eq:Main}
\frac{\langle Q^\text{tran} \rangle_\text{min}}{k_\text{B} T \ln2}
  & \equiv \lim_{N \rightarrow \infty}
  \left[ \frac{\langle Q_{0 \rightarrow N} \rangle_\text{min}}{k_\text{B} T
  \ln{2}} + N (\hmu' - \hmu) \right]
   \nonumber \\
  & = - \EE' + \I[\overleftarrow{Y}';\overrightarrow{Y}]
    + \I[X_0;\overleftarrow{Y}',\overrightarrow{Y}]
	~,
\end{align}
where $\EE' = \I[\overleftarrow{Y}';\overrightarrow{Y}']$ is the output
sequence's \emph{excess entropy} \cite{Crut01a}, $\I[A;B]$ is the mutual
information between random variables $A$ and $B$, $\overleftarrow{Y}$
($\overleftarrow{Y}'$) is the input (output) past---the sequence of input
(output) symbols that have already interacted with (been produced by) the
transducer, $\overrightarrow{Y}$ ($\overrightarrow{Y}'$) is the input (output)
future---the sequence of input (output) symbols that have not yet interacted
with (been produced by) the transducer, and $X_0$ is the random variable for
the transducer's state after sufficiently long time, such that
$\overleftarrow{Y}'$ and $\overrightarrow{Y}$ are both effectively
semi-infinite chains of random variables. The expression comes from shifting to
the ratchet's reference frame, so that at time $N$ state $X_N$ becomes $X_0$
and the currently interacting tape symbol is relabeled $Y_0$, rather than
$Y_N$. (Equation (\ref{eq:Main}) is proved in the Supplementary Materials.)

From it we conclude that the minimum transient cost has three components.
However, they are subtly interdependent and so we cannot minimize them
piece-by-piece to maximize thermodynamic efficiency. For instance, the first
term in the transient cost is a benefit of having correlation between the
output past and output future, qualified by $\EE'$. Without further thought,
one infers that outputs that are more predictable from their past, given a
fixed entropy production rate, are easier to produce thermodynamically.
However, as we see below when analyzing process generation, the other terms
cancel this benefit, regardless of the output process. Perhaps
counterintuitively, the most important factor is the output's
intrinsic structure.

The remaining two terms in the transient cost are the cost due to correlations
between the input and the output, quantified by
$\I[\overleftarrow{Y}';\overrightarrow{Y}]$, and the cost due to correlations
between the transducer and the entire input-output sequence, quantified by
$\I[X_0;\overleftarrow{Y}',\overrightarrow{Y}]$. The last term, which through
$X_0$ depends explicitly on the transducer's structure, shows how different
implementations of the same computation change energetic requirements. Said
differently, we can alter transducer states as well as their interactions with
tape symbols, all the while preserving the computation---the joint-input output
distribution---and this \textit{only} affects the last term in
Eq.~(\ref{eq:Main}). For this reason, we call it the \emph{minimal
implementation energy cost} $Q^\text{impl}$ given a transducer:
\begin{align}
(k_\text{B} T \ln2)^{-1} \langle Q^\text{impl} \rangle_\text{min}
	= \I[X_0;\overleftarrow{Y}',\overrightarrow{Y}]
	~.
\label{eq:CostImpl}
\end{align}
This cost extends beyond that due to predictively generating an output process
\cite{Garn15} to any type of input-output transformation. Having identified
this cost, we can then find thermodynamically efficient ratchets by choosing
implementations with the smallest mutual information between the transducer's
state and the output past and the input future.

\paragraph{Generating Structured Patterns} Paralleling Ref. \cite{Boyd15a}, we now consider the thermodynamic
cost of generating a sequential pattern of output symbols from a sequence of
IID input symbols. Since the latter are uncorrelated and we restrict ourselves
to nonanticipatory transducers (i.e., transducers with no direct access to
future input \cite{Barn13a}), the input future is statistically independent of
both the current transducer state and the output past:
$\I[X_0;\overleftarrow{Y}',\overrightarrow{Y}] = \I[X_0;\overleftarrow{Y}']$
and $\I[\overleftarrow{Y}';\overrightarrow{Y}] = 0$. As a result, we have the
following simplifications for the minimal transient dissipation and
implementation costs:
\begin{align}
\label{eq:TranGen}
(k_\text{B} T \ln2 )^{-1} \langle Q^\text{tran}
  \rangle_\text{min} & = \I[X_0;\overleftarrow{Y}'] - \EE^\prime  \\
  (k_\text{B} T \ln2 )^{-1} \langle Q^\text{impl}
  \rangle_\text{min} & = \I[X_0;\overleftarrow{Y}']
  \label{eq:ImplGen}
  ~.
\end{align}

\begin{figure}[tpb]
\centering
\includegraphics[width=\columnwidth]{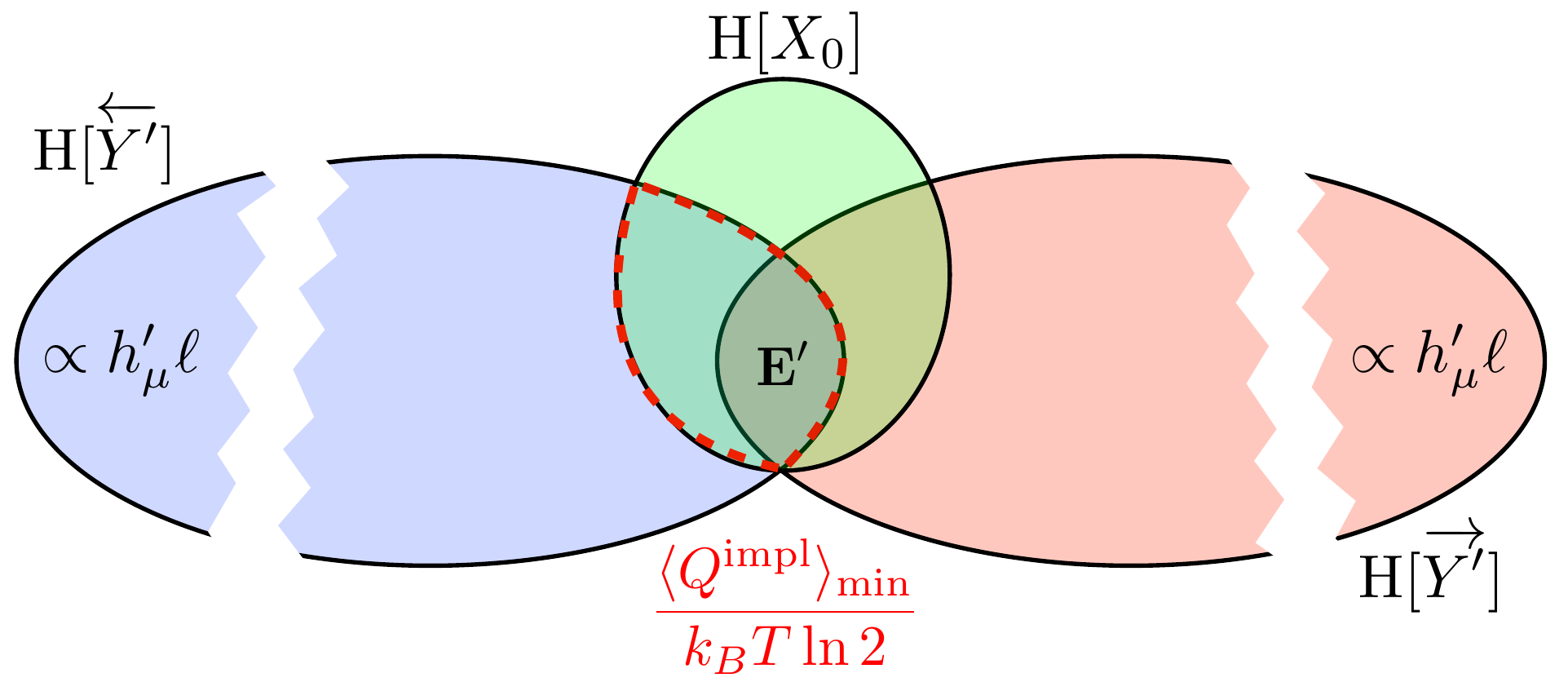}
\caption{Shannon measures for physical information transduction---general
	case of nonunifilar transducers: Transducer output past
	$\protect\overleftarrow{Y^\prime}$ and output future
	$\protect\overrightarrow{Y^\prime}$ left (blue) and right (red) ellipses, respectively;
	shown broken since the future and past entropies
	$\H[\protect\overleftarrow{Y^\prime}]$ and
	$\H[\protect\overrightarrow{Y^\prime}]$ diverge as $\hmu \ell$, with $\ell$
	being the length of past or future, respectively. $\H[X_0]$ illustrates the
	most general relationship the generating transducer state $X_0$ must have
	with the process future and past. Implementation cost
	$\I[X_0;\protect\overleftarrow{Y}']=\langle Q^\text{impl}
	\rangle_\text{min}/k_B T \ln 2 $ is highlighted by a dashed (red) outline.
	}
\label{fig:NonunifilarGeneratorIDiagram}
\end{figure}

The fact that the input is IID also tells us that the transducer's states are
also the internal states of the hidden Markov model (HMM) generator of the
output process \cite{Barn13a, Boyd15a}. This means that the transducer variable
$X_0$ must contain all information shared between the output's past
$\overleftarrow{Y}'$ and future $\overrightarrow{Y}'$ \cite{Crut08a, Crut01a},
as shown in the \textit{information diagram} in Fig.
\ref{fig:NonunifilarGeneratorIDiagram}. (Graphically, the $\EE^\prime$ atom is
entirely contained within $\H[X_0]$.) There, an ellipse depicts a variable's
Shannon entropy, an intersection of two ellipses denotes the mutual information
between variables, and the exclusive portion of an ellipse denotes a variable's
conditional entropy. For example, $\EE^\prime =
\I[\overleftarrow{Y}';\overrightarrow{Y}']$ is the intersection of
$\H[\overleftarrow{Y}']$ and $\H[\overrightarrow{Y}']$. And, the leftmost
crescent in Fig.~\ref{fig:NonunifilarGeneratorIDiagram} is the conditional
Shannon entropy $\H[\overleftarrow{Y^\prime}|X_0]$ of the output past
$\overleftarrow{Y^\prime}$ conditioned on transducer state $X_0$.  The diagram
also notes that this information atom, which is in principle infinite, scales
as $\hmu \ell$, where $\ell$ is the sequence length.

As stated above, Fig.~\ref{fig:NonunifilarGeneratorIDiagram} also shows that
the ratchet state statistically shields past from future, since the
ratchet-state entropy $\H[X_0]$ (green ellipse) contains the information
$\EE^\prime$ shared between the output past and future (overlap between (left)
blue and right (red) ellipses). Thus, the implementation cost
$\I[X_0;\overleftarrow{Y}^\prime]$, highlighted by dashed (red) outline,
necessarily contains the mutual information between the past and future. We are
now ready to find the most efficient thermodynamic implementations for a given
computation.

Both the asymptotic and transient bounds are achievable for the task of
generating a given process, as shown by Ref. \cite{Garn15}, through an
alternating sequence of adiabatic then quasistatic control of energy levels.
Thus, when we find an implementation that minimizes the bound on energy cost,
this also tells us the exact form of a physical device that implements the
transducer and achieves the bound.

\newcommand{\PS}{\AlternateState}
\newcommand{\ps}{\alternatestate}
\newcommand{\CS}{\CausalState}
\newcommand{\cs}{\causalstate}

\begin{figure}[tpb]
\centering
\includegraphics[width=\columnwidth]{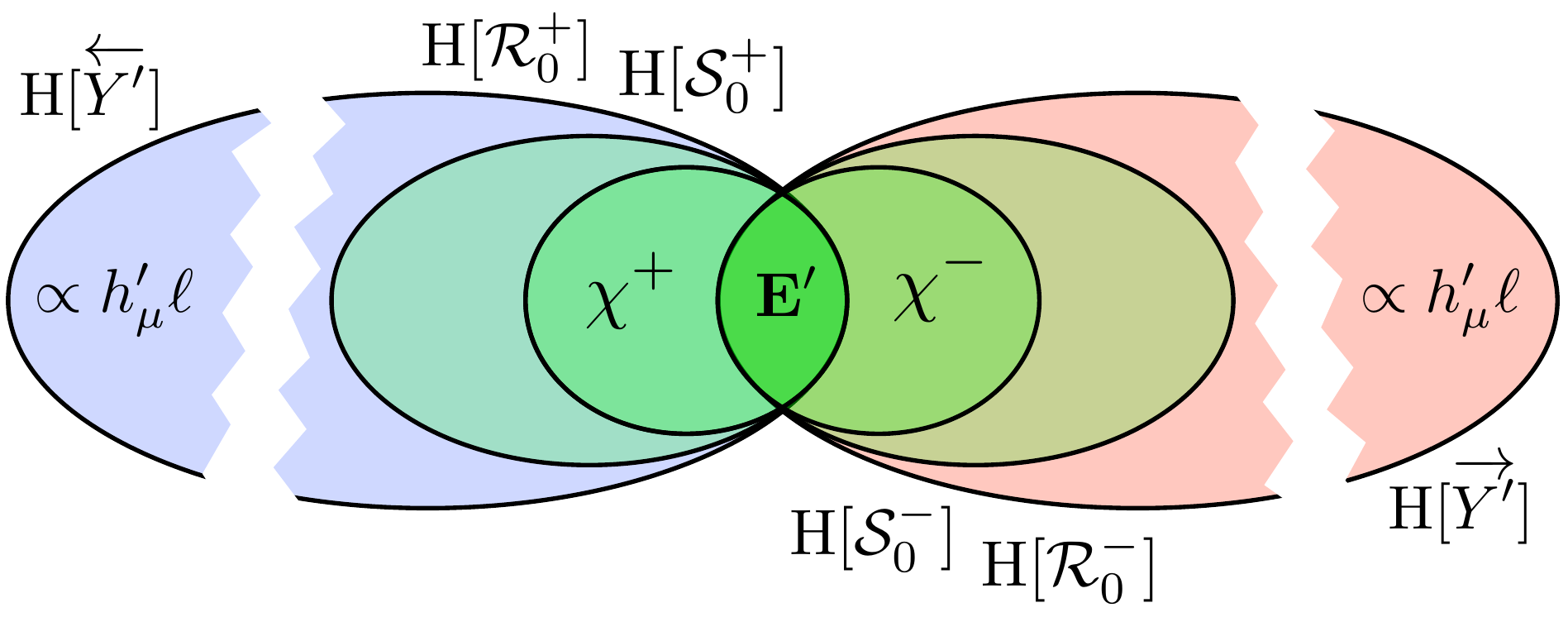}
\caption{Optimal physical information transducers---predictive and
	retrodictive process generators: Process generator variables, predictive 
	states $\PS$ and causal states $\CS$, denoted with green ellipses with
	latter contained within the former, being the minimal set of
	predictive states. A given process has alternative unifilar ($\PS^+_0$ or
	$\CS^+_0$) and co-unifilar generators ($\PS^-_0$ or $\CS^-_0$).
	Component areas are the sigma-algebra \emph{atoms}: conditional
	entropies---entropy rate $\hmu$ and crypticities $\PC^+$ and $\PC^-$---and
	a mutual information---the excess entropy $\EE^\prime$. Since the state
	random variables $\PS^+_0$ and $\CS^+_0$ are functions of the output past
	$\protect\overleftarrow{Y^\prime}$, their entropies are wholly contained
	within the past entropy $\H[\protect\overleftarrow{Y^\prime}]$. Similarly,
	co-unifilar generators, denoted by the random variables $\PS^-_0$ and
	$\CS^-_0$, are functions of output future
	$\protect\overrightarrow{Y^\prime}$. Thus, their entropies are contained
	within the output future entropy $\H[\protect\overrightarrow{Y^\prime}]$.
	The \eM\ generator with causal states $\CS^+_0$ is the unifilar generator
	with minimal Shannon entropy (area). The random variable $\PS^-_0$ realizes
	the current state of the minimal co-unifilar generator, which is the time
	reversal of the \eM\ for the time-reversed process \protect\cite{Elli11a}.
	Transducers taking the form of any of these generators produce the same
	process, but structurally distinct generators exhibit different
	dissipations and thermodynamic implementation costs.
	}
\label{fig:TransducerIDiagram}
\end{figure}

Consider first the class of predictive, \emph{unifilar} information
transducers; denote their states $\PS^+_0$. Unifilarity here says that the
current state $\PS^+_0$ is restricted to be a function of the semi-infinite
output past: the ratchet's next state $\PS^+_0$ is unambiguously determined by
$\overleftarrow{Y}^\prime$.

A unifilar information transducer corresponds to the case where the transducer
state entropy $\H[X_0 = \PS^+_0]$ has no area outside that of the output
past's entropy $\H[\overleftarrow{Y}']$. (See
Fig.~\ref{fig:TransducerIDiagram}.) As evident there, the implementation cost
$\I[X_0;\overleftarrow{Y}']$ is the same as the transducer's \emph{state
uncertainty}---the Shannon entropy $\H[X_0 = \PS^+_0]$. Thus, according to
Eq.~(\ref{eq:ImplGen}) the thermodynamically most efficient unifilar
transducer is that with minimal state-uncertainty $\H[X_0 = \CS^+_0]$---the
entropy of the \emph{\eM\ causal states} $\CS^+_0$ of computational mechanics
\cite{Crut12a}, which comprise the minimal set of predictive states
\footnote{One might conclude that simpler (smaller forward causal-state
entropy) is thermodynamically better (more efficient) \cite{Garn15}. Our
development shows when this holds and when not, leading to a broader and more
incisive view of optimal adaptive systems.}. This confirms the result that, if
one is restricted to \emph{predictive} generators, simpler is better
\cite{Garn15}.

There are further connections with computational mechanics. For \eM\
information transducers with causal states $\CS^+_0$, the mutual information
between the transducer and the output past is the output process' statistical
complexity: $\I[\CS^+_0;\overleftarrow{Y}'] = \Cmu^\prime$. In other words, the
minimal implementation cost of a pattern generated by a unifilar information
transducer is the pattern's statistical complexity. The transient dissipation
that occurs when generating a structured pattern, given in
Eq.~(\ref{eq:TranGen}), is then the output's crypticity $\PC^+ = 
\Cmu^\prime - \EE^\prime$ \cite{Crut08a}, as Ref. \cite{Elli11a} concluded previously and Ref. \cite{Garn15} more recently.

Now, consider the more general case in which we allow the transducer
implementation to be nonunifilar; see
Fig.~\ref{fig:NonunifilarGeneratorIDiagram} again. From the Data Processing Inequality \cite{Cove06a}, it follows that the mutual information between $X_0$ and $\overleftarrow{Y}'$ cannot be less than the output's excess entropy: 
\begin{align}
\I[X_0;\overleftarrow{Y}^\prime] \geq \EE^\prime
  ~.
\label{eq:Inequality}
\end{align} 
Thus, the minimum structural cost over alternate pattern-generator
implementations is therefore the output pattern's excess entropy.

Figure~\ref{fig:NonunifilarGeneratorIDiagram} suggests how to find this
minimum. The implementation cost highlighted by the dashed (red) line can be
minimized by choosing a transducer whose states are strictly functions of the
future. In this case, the transducer's mutual information with the output past
is simply $\EE^\prime$, achieving the bound on implementation cost given by Eq.
(\ref{eq:Inequality}). (Refer now to Fig.~\ref{fig:TransducerIDiagram}.)
Constructed using states that are functions of the future, such a ratchet is a
generator with retrodictive (as opposed to predictive) states, denoted
$\PS^-_0$ or $\CS^-_0$ \cite{Crut08b}. This means that the generator is
\emph{counifilar}, as opposed to unifilar \cite{Maho11a, Crut10a}.  These
generators have the same states as the unifilar generators of the time-reversed
process.  Retrodictive generators produce the same output process by running
along the information reservoir in the same way as the predictive generators,
but rather than store all of the information in the past outputs required to
\emph{predict} the future, they only store just enough to \emph{generate} it.
This affords them a fundamental energetic advantage.

Critically, any such retrodictive implementation is maximally efficient,
dissipating zero transient heat $\langle Q^\text{tran} \rangle_\text{min}=0$,
even though the state uncertainty varies across implementations: $\H[\PS^-_0] >
\H[\CS^-_0]$. Unlike unifilar transducers, for a given output process there are
infinitely many counifilar information transducers of varying state-complexity
that are all maximally thermodynamically efficient. In other words, simpler is
not necessarily thermodynamically better for optimized transducers. This shows,
as a practical matter, that both the design and evolution of efficient
biological computations have a wide latitude when it comes to physical
instantiations.

To summarize, we identified the transient and structural thermodynamic costs of
physical information transduction, generalizing the recent Information Process
Second Law. These bound the energetic costs incurred by any physically embedded
adaptive system as it comes synchronize with the states of a structured
environment. When asking about which physical implementations are the most
thermodynamically efficient we showed that they are retrodictive generators,
not necessarily \eMs.

\paragraph*{Supplementary Materials:} Derivations and further
discussion and interpretation.

\acknowledgments

As External Faculty, JPC thanks the Santa Fe Institute for its hospitality
during visits. This work was supported in part by the U. S. Army Research
Laboratory and the U. S. Army Research Office under contract W911NF-13-1-0390.


\bibliography{chaos}

\newpage
\clearpage

\appendix

\begin{center}
\large{Supplementary Materials}\\
\normalsize
for\\
\emph{Transient Dissipation and Structural Costs of\\
Physical Information Transduction}\\
Alexander B. Boyd, Dibyendu Mandal, Paul M. Riechers, and James P. Crutchfield
\end{center}

\setcounter{equation}{0}
\setcounter{figure}{0}
\setcounter{table}{0}
\setcounter{page}{1}
\makeatletter
\renewcommand{\theequation}{S\arabic{equation}}
\renewcommand{\thefigure}{S\arabic{figure}}
\renewcommand{\bibnumfmt}[1]{[S#1]}

\section{Shannon versus Kolmogorov-Sinai Entropy Rates}

On the one hand, it is now commonplace shorthand in physics to describe a
symbol-based process in terms of Shannon's information theory and so measure
its intrinsic randomness via the Shannon entropy rate. On the other, properly
capturing the information processing behavior of physically embedded systems is
more subtle. There are key conceptual problems. For one, the symbol-based
Shannon entropy rate may not be well defined for a given continuous physical
system. In the present setting we consider the entire engine as a physical
system. Then, $\hmu$ and $\hmu^\prime$ are the Kolmogorov-Sinai entropies of
the associated reservoir dynamical systems [S1,S2]. They are well defined
suprema over all coarse-grainings of the system's state space.

\section{Information Reservoirs Beyond Tapes}

Rather than implement the information reservoir as a tape of symbols, one can
simply employ a one-dimensional lattice of Ising spins. Moreover, the reservoir
need \emph{not} be 1D, but this is easiest to analyze since the total entropy
of a 1D sequence is related to (but not equal to) the Shannon entropy rate and
an engine simply accesses information by moving sequentially along the tape.
Higher-dimension reservoirs, even with nonregular topologies connecting the
information-bearing degrees-of-freedom, can be a thermodynamic resource when
there is large total correlation among its degrees of freedom, at the cost of
decorrelating the information-bearing degrees of freedom [S3, Ch. 9].

\section{Thermodynamics of General Computation}

We focused on spatially-unidirectional information transduction due to the
stronger thermodynamic results. However, the thermodynamic results are valid
much more broadly, applicable to Turing-equivalent machines as well as non-1D
information transduction, as just noted.

First, Turing machines that move unidirectionally, reading input tape cells
once and writing results only once to an output tape, are equivalent to the
transducers used here. However, unidirectional Turing machines employ internal
tapes as scratch storage [S4]
and this now-internal memory must be
taken into account when assessing thermodynamic resources.

Second, the choice of \emph{implementation} of a particular computation implies
a transient thermodynamic cost above the asymptotic implementation-independent
work rate. The general result is:
\begin{align*}
\langle Q^\text{tran}_{0 \rightarrow N}\rangle / k_\text{B} T \ln{2}
  \geq \H[X_0, \mathbf{Y}_0] - \H[X_N, \mathbf{Y}_N] + N \Delta h
  ~,
\end{align*}
where $\mathbf{Y}_N$ is the random variable for the information-bearing degrees of
freedom at time $N$ and $\Delta h$ is the difference in the extensive component
of the entropy density of the output and input tape processes. In short,
\emph{the transient cost due to an implementation stems from the correlation
built up between the device's state and the pattern on which it acts,
discounted by the intensive part of the output pattern's entropy.}

\section{Origin of Transient Information Processing Costs}

We demonstrate how the transient IPSL of Eq.~(\ref{eq:Main}) arises. The steps
give additional insight.

Assuming that we are able to achieve asymptotic IPSL bounds---say, as in Ref.
\cite{Garn15}---the cumulative transient cost of information processing over the
interval $t \in [0,N]$ is given by:
\begin{align}
\langle Q^\text{tran}_{0 \rightarrow N} \rangle
  \equiv \langle Q_{0 \rightarrow N} \rangle - N (h_\mu - h'_\mu) k_\text{B} T \ln{2}
  ~. 
\label{eq:TranDef}
\end{align} 
Combining with Eq.~(\ref{eq:FullBound}), yields:
\begin{align*}
& \frac{\langle Q^\text{tran}_{0 \rightarrow N}\rangle}{k_\text{B} T \ln{2}}
  \geq \H[X_0, \boldsymbol{Y}_0] - \H[X_N, \boldsymbol{Y}_N] + N (\hmu^\prime - \hmu) \\
  & \quad = \H[X_0, Y_{0:\infty}] - \H[X_N, Y'_{0:N}, Y_{N, \infty}]
  	+ N(\hmu^\prime - \hmu) \\
  & \quad = (\H[X_0] + \H[Y_{0:\infty}] - \I[X_0;Y_{0:\infty}]) \nonumber \\
	& \qquad -(\H[X_N] + \H[Y'_{0:N}, Y_{N:\infty}] - \I[X_N; Y'_{0:N}, Y_{N:\infty}]) \\
	& \qquad + N(\hmu^\prime - \hmu)
  ~.
\end{align*}
The last line used the standard identify $\H[A,B] = \H[A] + \H[B] - \I[A;B]$
for random variables $A$ and $B$. Since we are interested in the purely
transient cost and not spurious costs arising from arbitrary initial
conditions, we start the engine in its stationary state, resulting in
stationary behavior, so that $\H[X_0]$ is the same as $\H[X_N]$. Furthermore,
we assume that the engine's initial state is uncorrelated with the incoming
symbols and so disregard $\I[X_0;Y_{0:\infty}]$. We then decompose the terms
$\H[Y'_{0:N}, Y_{N:\infty}]$ and $\H[Y_{0:\infty}] \equiv \H[Y_{0:N},
Y_{N:\infty}]$ according to above. These assumptions and decompositions lead to:
\begin{align}
\frac{\langle Q^\text{tran}_{0 \rightarrow N}\rangle}{k_\text{B} T \ln{2}}
  & \geq \H[Y_{0:N}] + \H[Y_{N:\infty}] - \I[Y_{0:N};Y_{N:\infty}] \nonumber \\
  & \qquad - \H[Y'_{0:N}] - \H[Y_{N:\infty}] + \I[Y'_{0:N};Y_{N:\infty}]
  \nonumber \\
  & \qquad + \I[X_N; Y'_{0:N}, Y_{N:\infty}] + N (\hmu^\prime - \hmu)
  ~.
\label{eq:TranProof2}
\end{align}

In the limit of large $N$, in which the transducer has interacted with a
sufficiently large number of input symbols, we can invoke the following
definitions of excess entropy:
\begin{align*}
\EE & = \lim_{N\rightarrow \infty}
  (\H[Y_{0:N}] - N h_\mu) \\
  & = \lim_{N \rightarrow \infty} I[Y_{0:N}; Y_{N:\infty}] \\
  \EE^\prime & = \lim_{N\rightarrow \infty} (H[Y'_{0:N}] - N h'_\mu)
  ~. 
\end{align*}
Upon shifting to the ratchet's reference frame and switching back to the more
intuitive notation: $X_N \rightarrow X_0$, $Y_{0:N} \rightarrow
\overleftarrow{Y}$, $Y'_{0:N} \rightarrow \overleftarrow{Y}'$, and $
Y_{N:\infty} \rightarrow \overrightarrow{Y}$, in which a left arrow means the
past and a right arrow the future, and invoking the above definitions, 
new notation and these definitions, we rewrite the inequality
Eq.~(\ref{eq:TranProof2}), after some cancellation, as:
\begin{align*}
\frac{\langle Q^\text{tran}\rangle}{k_\text{B} T \ln{2}}
  & \geq - \EE^\prime + \I[\overleftarrow{Y}';\overrightarrow{Y}]
  + \I[X_0; \overleftarrow{Y'}, \overrightarrow{Y}]
  ~,
\end{align*}
where $\langle Q^\text{tran} \rangle$ is the total transient cost over infinite
time, $\langle Q^\text{tran} \rangle = \lim_{N \rightarrow \infty} \langle Q_{0
\rightarrow N}^\text{tran}\rangle$. This is our main result,
Eq.~(\ref{eq:Main}), and the starting point for the other results.

\section*{References}

\begin{enumerate}
\item[[S1\!\!]]
A.~N. Kolmogorov.
\newblock Entropy per unit time as a metric invariant of automorphisms.
\newblock {\em Dokl. Akad. Nauk. SSSR}, 124:754, 1959.
\newblock (Russian) Math. Rev. vol. 21, no. 2035b.

\item[[S2\!\!]]
Ja.~G. Sinai.
\newblock On the notion of entropy of a dynamical system.
\newblock {\em Dokl. Akad. Nauk. SSSR}, 124:768, 1959.

\item[[S3\!\!]]
P.~M. Riechers.
\newblock {\em Exact Results Regarding the Physics of Complex Systems via
  Linear Algebra, Hidden Markov Models, and Information Theory}.
\newblock PhD thesis, University of California, Davis, 2016.

\item[[S4\!\!]]
J.~E. Hopcroft, R.~Motwani, and J.~D. Ullman.
\newblock {\em Introduction to Automata Theory, Languages, and Computation}.
\newblock Prentice-Hall, New York, third edition, 2006.
\end{enumerate}

\end{document}